\documentstyle[epsfig]{article}
\begin{document}

\title{Low-energy electronic structure in Y$_{1-x}$Ca$_{x}$Ba$_{2}$Cu$_{3}$O$%
_{7-\delta }:$ comparison of time-resolved optical spectroscopy, NMR,
neutron\ and tunneling data.}
\author{J.Demsar, K.Zagar, V.V.Kabanov and D.Mihailovic \\
%EndAName
''J.Stefan Institute'', Jamova 39, 1001 Ljubljana, Slovenia}
\maketitle

\begin{abstract}
Time-resolved optical measurements give information on the quasiparticle
relaxation dynamics in YBCO, from which the evolution of the gap with doping
and temperature can be systematically deduced. In this paper these optical
charge-channel ''pseudogap'' data are compared with the ''pseudogap''
obtained from the NMR Knight shift K$_{s},$ spin polarized neutron
scattering (SPNS) and single particle tunneling measurements. A simple
energy level diagram is proposed to explain the different ''gap'' magnitudes
observed by different spectroscopies in YBa$_{2}$Cu$_{3}$O$_{7-\delta }$,
whereby the spin gap $\Delta _{s}$ in NMR and SPNS corresponds to a triplet
local pair state, while $\Delta _{p}$ in the charge excitation spectrum
corresponds to the pair dissociation energy. At optimum doping and in the
overdoped state, an additional $T$-dependent gap becomes evident, which
closes at $T_{c}$, suggesting a cross-over to a more conventional BCS-like
superconductivity scenario.
\end{abstract}

\section{Introduction}

Spectroscopic studies of cuprates over the years have shown that at low
energies these materials exhibit quite a complex spectral structure, which
changes with temperature and doping in a complicated way. Often there
appears to be reasonable agreement regarding some of the main features
between various experimental techniques. For example, optical femtosecond
quasiparticle relaxation measurements and single-particle (Giever) tunneling
show a similar size pseudogap in the spectrum over a large portion of the
phase diagram, and the latter is remarkably similar to the spectral features
measured by angle-resolved photoemission. However, the pseudogap $\Delta
_{s} $ as observed by spectroscopies like NMR\ and spin-polarized neutron
scattering appears to be smaller than the optical and tunneling pseudogap $%
\Delta _{p}$ by approximately a factor of 2, for which there is as yet no
accepted theoretical explanation. Thus in spite of the availability of
spectral data over a large range of doping in many materials, the
interpretatation of the low-energy excitation spectrum is still highly
controversial. In this paper we summarize some of the results of
time-resolved quasiparticle recombination spectroscopy as a function of
doping and temperature in YBa$_{2}$Cu$_{3}$O$_{7-\delta }$, which appear to
give qualitative new insight into the origin of the low-energy spectral
features of this material and its phase diagram. We compare the
doping-evolution of the pseudogap $\Delta _{p}$ obtained from time-resolved
spectroscopy with a quantitative analysis of the ''spin-gap'' from the NMR
knight shift $K_{s}$ in the underdoped state using the same type of $T$%
-independent gap as deduced from the quasiparticle relaxation data. We find
that the time-resolved QP relaxation data are quite consistent with Giever
tunneling and ARPES data and suggest that all the observations together can
be explained by the existence of a pair-breaking pseudogap $\Delta _{p}$ and
a triplet-state local pair excitation $\Delta _{s}$ at $E\approx \Delta
_{p}/2$.

\section{Time-resolved spectroscopy results}

The time-resolved optical spectroscopy applied to superconductors and other
materials with a gap has been given in detail elsewhere\cite
{Kabanov,DemsarPRL}, so here we shall discuss only the results. The
equations describing the temperature dependence of photoexcited QP density
and their lifetime as a function of temperature are given by Kabanov et al
\cite{Kabanov}. Using these equations, the properties of the gap can be
investigated as a function of doping and temperature.
%TCIMACRO{
%\TeXButton{TeX field}{\begin{figure} [hb]
%\centerline{\epsfig{file=figure1.eps,height=3in}}
%\caption{The photoinduced reflectivity in
%YBa$_{2}$Cu$_{3}$O$_{6.95}$ showing the two-component relaxation below
%$T_{c} $ attributed to the two gaps $\Delta_{p}$ and $\Delta_{BCS}$. }
%\end{figure} %
%}}%
%BeginExpansion
\begin{figure} [hb]
\centerline{\epsfig{file=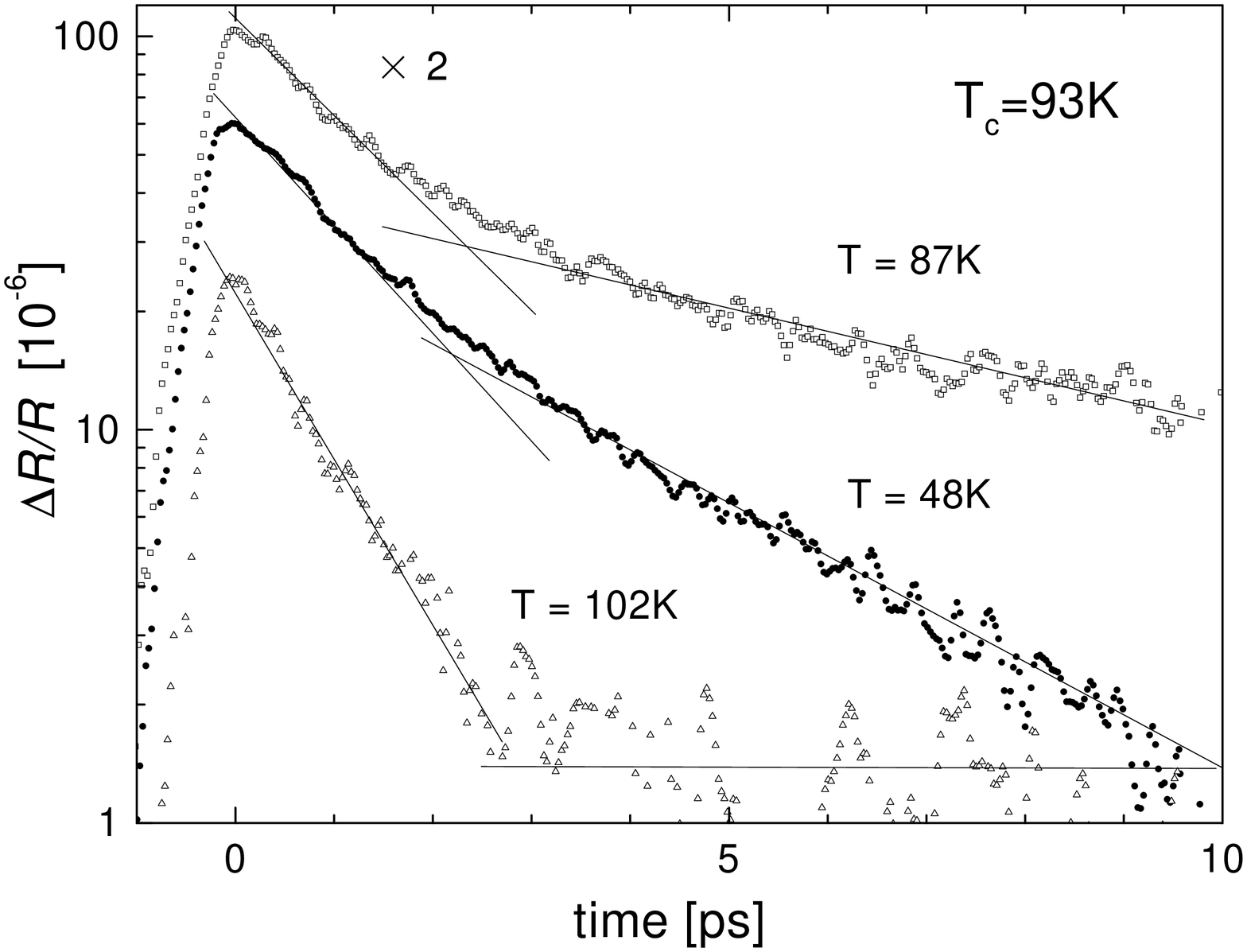,height=3in}}
\caption{The photoinduced reflectivity in
YBa$_{2}$Cu$_{3}$O$_{6.95}$ showing the two-component relaxation below
$T_{c} $ attributed to the two gaps $\Delta _{p}$ and $\Delta _{BCS}$. }
\end{figure} %
%
%EndExpansion
Systematic experiments on YBaCuO\ as a function of O concentration over a
wide range of $\delta $ have shown very clear and systematic behavior of the
QP dynamics \cite{Kabanov}. Particularly important is that the results are
quite insensitive to surface quality and have been repeated after a period
of a year with the same results. Single crystals were also studied and no
significant differences were found in comparison with the thin film data.
These measurements have recently been extended to the overdoped state using
calcium-doped (Y,Ca)Ba$_{2}$Cu$_{3}$O$_{7-\delta }$ crystals by Demsar et al
\cite{DemsarPRL}, giving data on the whole phase diagram from insulating and
underdoped to overdoped YBCO. A typical time-resolved signal in the
optimally doped phase is shown in Figure 1. Above $T_{c}$, one exponential
short-lived decay is observed whose amplitude is decreasing as $T$ is
increased (Figure 2), while below $T_{c}$, two exponentials are clearly
seen, one of which has a longer lifetime which is temperature-\textit{%
dependent }and diverges as $T\rightarrow T_{c}$. Such behaviour is typical
also for overdoped samples, but not for underdoped samples, where only a
single exponential decay is observed, with a $T$-indendent time constant $%
\tau _{p}\sim 0.4$ ps showing no anomaly at $T_{c}$. From these experiments
we can deduce that in underdoped state the evolution of the QP dynamics with
temperature and doping is dominated by a temperature-independent pseudogap $%
\Delta _{p}$. Near optimum doping an \textit{additional} BCS-like
temperature-dependent gap $\Delta _{BCS}$ appears below $T_{c},$ which is
present also in all the overdoped samples \cite{DemsarPRL} and gives rise to
the additional temperature-dependent relaxation process below $T_{c}$.
%TCIMACRO{
%\TeXButton{TeX field}{\begin{figure}[hb]
%\centerline{\epsfig{file=figure2.eps,height=3.5in}}
%\caption{The amplitude of the photoinduced absorption (PIA)\ as a function of
%temperature for optimally doped YBa$_{2}$Cu$_{3}$O$_{7-\delta }$ ($\delta
%\sim 0.05$). Overdoped samples with substituted Ca for Y show similar
%2-component $T$-dependence for all Ca concentrations. b) The temperature
%dependence of the PIA for underdoped YBa$_{2}$Cu$_{3}$O$_{7-\delta }$
%($\delta >0.15$). The lines are the theoretical fits to the data [Ref. 1]
%with
%$\Delta _{p}$ or $\Delta _{BCS}(T=0K)$ as the only fitting parameter.}
%\end{figure}
%
%}}%
%BeginExpansion
\begin{figure}[hb]
\centerline{\epsfig{file=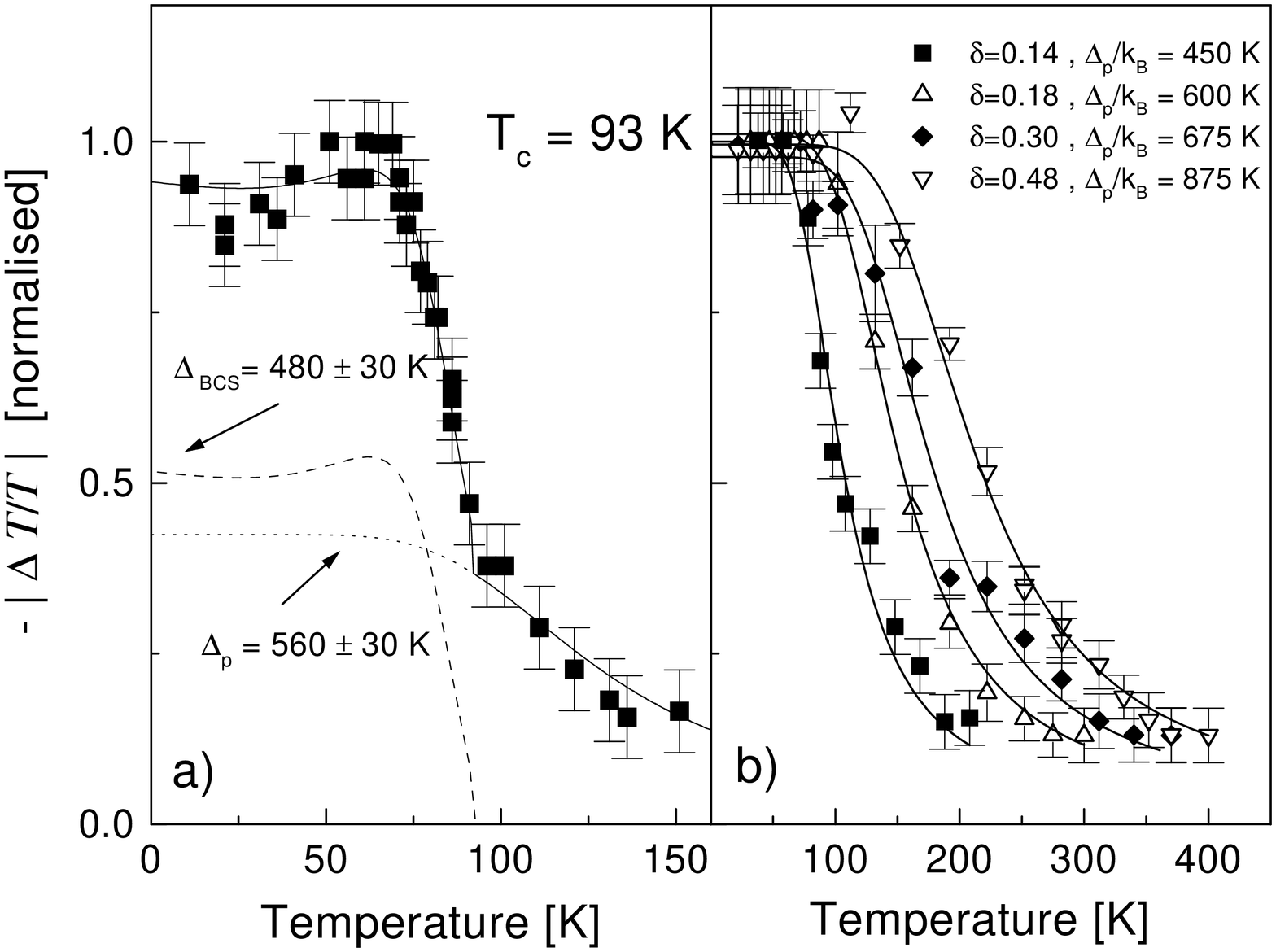,height=3.5in}}
\caption{The amplitude of the photoinduced absorption (PIA)\ as a function of
temperature for optimally doped YBa$_{2}$Cu$_{3}$O$_{7-\delta }$ ($\delta
\sim 0.05$). Overdoped samples with substituted Ca for Y show similar
2-component $T$-dependence for all Ca concentrations. b) The temperature
dependence of the PIA for underdoped YBa$_{2}$Cu$_{3}$O$_{7-\delta }$
($\delta >0.15$). The lines are the theoretical fits to the data [Ref. 1]
with
$\Delta _{p}$ or $\Delta _{BCS}(T=0K)$ as the only fitting parameter.}
\end{figure}

%
%EndExpansion

The fact the $\tau $ diverges near $T_{c}$ unambiguously signifies that $%
\Delta _{BCS}\rightarrow 0$ at $T_{c}$. Thus close to optimum doping and in
the overdoped state two gaps appear to be present simultaneously, a feature
consistent with the spatially inhomogeneous ground state\cite
{MihailovicMuller} in which the pair-breaking excitation is represented by $%
\Delta _{p}$, while the $\Delta _{BCS}$ is a gap associated with the
collective behaviour of high-carrier density stripes or clusters which start
to form near optimum doping. The values of the gaps $\Delta _{p}$ and $%
\Delta _{BCS}$ as a function of doping obtained by fitting the temperature
dependence of the photoinduced signal amplitude as shown in Fig. 2 with the
model calculation of Kabanov et al \cite{Kabanov} are plotted in Figure 4.

\section{Comparison with other spectroscopies}

From the time-resolved data on the underdoped YBCO we have deduced that the
low-energy spectrum can be described by a\emph{\ }single\emph{\
temperature-independent gap} $\Delta _{p}$. Starting from this observation
we decided to analyse the temperature dependence of the NMR\ Knight shift $%
K_{s}$ available from the literature using the same $T$-indepedent $\Delta
_{p.}$ The aim is a) to see if the simple model can describe the $T$%
-dependence of K$_{s}$ and b) to see if the values of the pseudogap obtained
for the NMR K$_{s}$ agree with the optical values. The NMR\ knight shift for
such a case can be written as \cite{AKM} $K_{s}=K_{0}+AT^{-\beta }\exp
[-\Delta _{s}/k_{B}T],$ where $K_{0}$ is the value of $K_{s}$ at zero
temperature, $\beta $ is an exponent which depends on the shape of the
singularity of the DOS at the gap and $A$ is a constant depending on the NMR
nucleus.
%TCIMACRO{
%\TeXButton{TeX field}{\begin{figure}[hb]
%\centerline{\epsfig{file=figure3.eps,height=3.5in}}
%\caption{The $^{89}$Y NMR\
%Knight shift K$_{s}$ as a function of temperature in
%YBa$_{2}$Cu$_{3}$O$_{7-\delta }$ for different $\delta
%$\protect\cite{Alloul}. b) K$_{s}$ as a
%function of temperature in YBa$_{2}$Cu$_{4}$O$_{8}$ ($T_{c}$=81 K). The
%solid squares data are for $^{63}$Cu from Curro et al \protect\cite{Curro}
%and the open circles for $^{17}$O from Williams et al \protect\cite{Williams}.}
%\end{figure}
%
%}}%
%BeginExpansion
\begin{figure}[hb]
\centerline{\epsfig{file=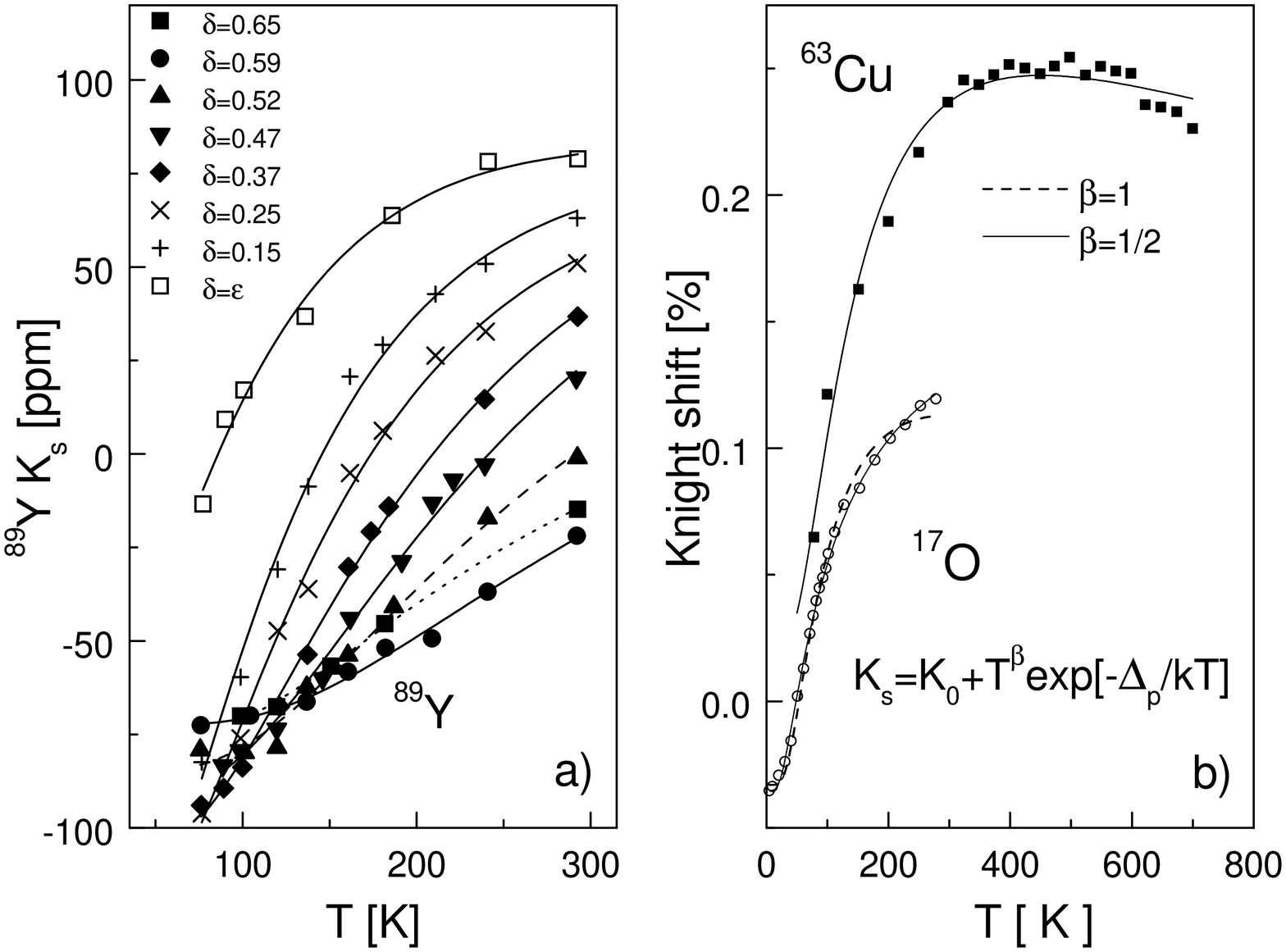,height=3.5in}}
\caption{The $^{89}$Y NMR\
Knight shift K$_{s}$ as a function of temperature in
YBa$_{2}$Cu$_{3}$O$_{7-\delta }$ for different $\delta
$\protect\cite{Alloul}. b) K$_{s}$ as a
function of temperature in YBa$_{2}$Cu$_{4}$O$_{8}$ ($T_{c}$=81 K). The
solid squares data are for $^{63}$Cu from Curro et al \protect\cite{Curro}
and the open circles for $^{17}$O from Williams et al \protect\cite{Williams}.}
\end{figure}

%
%EndExpansion
The results of fits to published data in YBCO 123 and 124 on $^{89}$Y\cite
{Alloul}, $^{63}$Cu \cite{Curro} and $^{17}$O\cite{Williams} are shown in
Figure 3 using $\beta =1/2.$ In spite of its simplicity, the model appears
to describe the data very well. The gap values $\Delta _{s}$ with $\beta
=1/2 $ obtained for YBa$_{2}$Cu$_{3}$O$_{7-\delta }$ are shown in Figure 4. $%
\Delta _{s}$ appears consistently lower than the $\Delta _{p}$ by
approximately a factor of 2.

Apart from NMR $K_{s}$, spin-polarized neutron scattering (SPNS) also shows
a spin-excitation peak at 34 meV ($\sim $390 K) in underdoped YBa$_{2}$Cu$%
_{3}$O$_{6.6}$, which is smaller than $\Delta _{p}$ by approximately a
factor of 2. However, the anomalies in the phonons signifying the presence
of \emph{charge }excitations occur near 70 meV ($\sim $810K), also
approximately twice the spin excitation energy and very close to the $\Delta
_{p}$ in Fig. 4. (see Mook et al, this volume).
%TCIMACRO{
%\TeXButton{TeX field}{\begin{figure}[hb]
%\centerline{\epsfig{file=figure4.eps,height=4.5in}}
%\caption{The energy gap(s) $\Delta _{p}$ and $\Delta _{s}$ and $\Delta
%_{BCS}$ as\ a function of doping. The open squares are from
%time-resolved QP relaxation \protect\cite{Kabanov,DemsarPRL}. The full
%circles are from NMR \protect\cite{Alloul}. The open diamonds are the
%neutron data (H.Mook this volume), while the full diamondsare the tunneling
%data (G.Deutscher, this volume). The solid squares represent $\Delta_{BCS}$
%from time-resolved data.}
%\end{figure}
%
%}}%
%BeginExpansion
\begin{figure}[hb]
\centerline{\epsfig{file=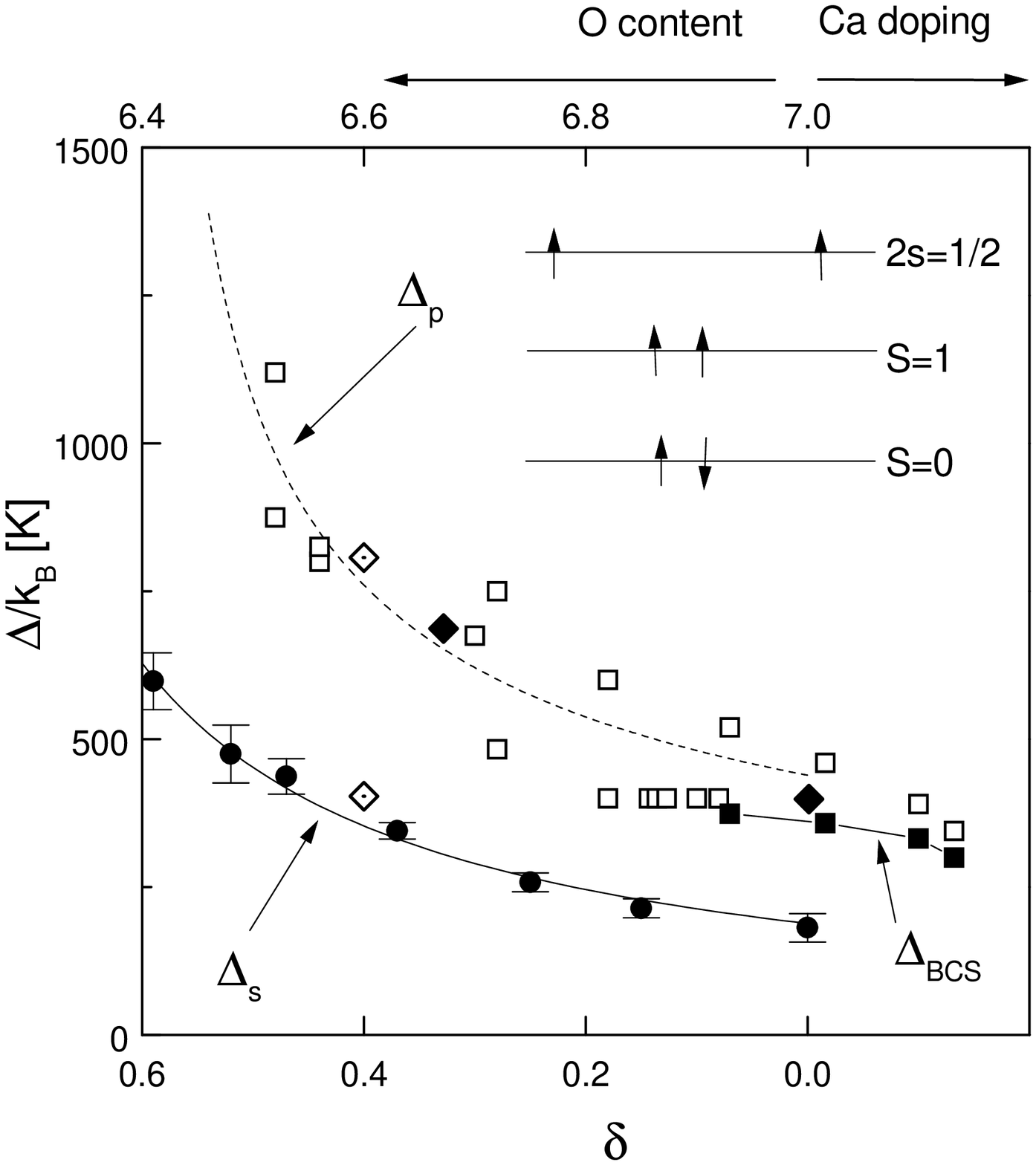,height=4.5in}}
\caption{The energy gap(s) $\Delta _{p}$ and $\Delta _{s}$ and $\Delta
_{BCS}$ as\ a function of doping. The open squares are from
time-resolved QP relaxation \protect\cite{Kabanov,DemsarPRL}. The full
circles are from NMR \protect\cite{Alloul}. The open diamonds are the
neutron data (H.Mook this volume), while the full diamonds are the tunneling
data (G.Deutscher, this volume). The solid squares represent $\Delta _{BCS}$
from time-resolved data.}
\end{figure}

%
%EndExpansion

To explain the two different gaps $\Delta _{p}$ and $\Delta _{s}$, we
propose a rather straightforward electronic structure in YBCO. A schematic
diagram is shown in the insert to Fig. 4. The ground state is composed of
\textit{local} S=0 singlet Cooper pairs. Since $\Delta _{p}$ is a QP\ charge
excitation it is clearly associated with pair breaking. However, if $%
J<\Delta _{p}$, the triplet local pair state also exists and lies within the
gap. It should be visible by spin-flip spectroscopies like NMR and SPNS, but
not by optical spectroscopy or SP\ tunneling which are charge excitations.
We therefore propose that the $\Delta _{s}$ observed by NMR\ and SPNS is the
S=1 \textit{local pair} triplet excited state. From the experimental data in
Fig. 4, both $\Delta _{p}$ and $\Delta _{s}$ decrease with increasing
doping, more or less as 1/$x$, where $x$ is the hole density. For low
doping, at $\delta \sim 0.6,$ $J$ $\approx $ 800K, consistent with Raman and
neutron measurements. We note that possibly the situation might be different
in La$_{2-x}$Sr$_{x}$CuO$_{4}$, where the optical gap and the NMR\ gap
appear to have the same energy scale\cite{Mueller}, suggesting that either 
$J$ > $\Delta _{p}$ in this material or that the triplet pair state is not
a bound state in this material.

In underdoped YBCO, at $T_{c}$ there is no anomaly in either the QP
relaxation\cite{Kabanov} or the NMR K$_{s}\cite{Alloul,Curro}$. Similarly
the SPNS intensity at 34 meV shows no anomaly at $T_{c}$ itself, but only a
gradual drop with increasing $T$. From this we can deduce that there is no
change in the DOS at $T_{c}$ in the underdoped state, and that all changes
of the DOS are gradual with $T$. This is equivalent to there being no
condensation energy associated with the superconducting transition itself
which is consistent with the Bose condensation of local pairs scenario,
where at $T_{c}$ macrosopic phase coherence is established with no change in
pairing amplitude. A 3D superconducting state forms when phase coherence
percolates through the entire sample resulting in a transition to a coherent
macroscopic state at $T_{c}$. In contrast, in optimally doped and overdoped
YBCO as the carrier density increases, the pairs begin to overlap,
collective effects become important and a transition to a more conventional
BCS-like scenario takes place\cite{Kabanov}. QP relaxation, NMR K$_{s}$,
SPNS and tunneling data all show an abrupt anomaly at $T_{c}$ signifying
that pairing and phase cohrence occur at the same (or nearly the same)
temperature.

The most important feature of the QP relaxation data not available from
other spectroscopies is the simultaneous unambiguous observation of 2 gaps
in optimally doped and overdoped samples, one \textit{T-independent} $\Delta
_{p}$ and the other \textit{T-dependent} $\Delta _{BCS}$ , with a BCS-like
\textit{T }dependence.

\end{document}